# Option Valuation using Fourier Space Time Stepping [*]

Kenneth R. Jackson[a], Sebastian Jaimungal[b] and Vladimir Surkov[a]

[a]Department of Computer Science, University of Toronto

[b]Department of Statistics, University of Toronto

It is well known that the Black-Scholes-Merton model suffers from several deficiencies. Jump-diffusion and Lévy models have been widely used to partially alleviate some of the biases inherent in this classical model. Unfortunately, the resulting pricing problem requires solving a more difficult partial-integro differential equation (PIDE) and although several approaches for solving the PIDE have been suggested in the literature, none are entirely satisfactory. All treat the integral and diffusive terms asymmetrically and are difficult to extend to higher dimensions. We present a new, efficient algorithm, based on transform methods, which symmetrically treats the diffusive and integrals terms, is applicable to a wide class of path-dependent options (such as Bermudan, barrier, and shout options) and options on multiple assets, and naturally extends to regime-switching Lévy models. We present a concise study of the precision and convergence properties of our algorithm for several classes of options and Lévy models and demonstrate that the algorithm is second-order in space and first-order in time for path-dependent options.

*The authors thank the Natural Sciences and Engineering Research Council of Canada for partially funding this work.

## 1. Introduction

The seminal works of Black and Scholes (1973) and Merton (1973) (BSM model) revolutionized our understanding of financial contracts with embedded options. Based on the assumption that stock prices follow a geometric Brownian motion, i.e. stock returns have a log-normal distribution, they demonstrated that a replicating strategy reduced the pricing problem to solving a partial differential equation (PDE) which is independent of the return of the asset. Today, option traders quote prices in terms of implied volatilities induced by matching market prices with those of the BSM model; however, it is well known that the BSM model suffers from several deficiencies rendering it inconsistent with market price behaviour. These inconsistencies manifest themselves in, for example, the observed implied volatility smile (or skew) and term structure. Various lines of research aim to remove these pricing biases by focusing on disparate extensions. One line of research seeks to introduce state dependence resulting in correlations between prices and volatility levels (see e.g. Derman and Kani (1994) and Duprie (1994) for nonparametric models and Cox and Ross (1976), Ingersoll (1997), and Rady (1997) for parametric models). Another line of research elevates volatility to a stochastic variable (e.g. Heston (1993)), or assumes volatility undergoes regime changes (e.g. Naik (1993)). A third line of research focuses on introducing jumps into the prices process itself (e.g. Merton (1976) and Madan and Seneta (1990)) while maintaining time homogeneity. All of these directions are able to correct for different aspects of the implied volatility surface and have their own unique set of advantages and disadvantages.

In this paper, we focus on pricing options where the underlying index, or indices, are driven by Lévy processes both with and without regime changes. This combines two of the three modeling directions and we succeed in developing an efficient method of pricing for a wide class of options. In all, there are four main purposes for this paper: first, to introduce a new numerical method based on the Fourier transform of the pricing partial integro-differential equation (PIDE); second, to study the algorithmic performance for various European and path-dependent options; third, to extend the method for multi-asset path-dependent contingent claims; and, fourth, to incorporate regime changes.

The option pricing problem under the BSM model can be reduced to solving a second-order parabolic PDE with the independent variables being time and stock price. By changing terminal or boundary conditions, or imposing early exercise constraints, the PDE can be used to price a variety of options. Under jump models, a PIDE with a non-local integral term must now be solved. A quick review of exponential Lévy models and the pricing PIDE is provided for completeness in section 2. An assortment of finite difference methods for solving these PIDEs have been proposed in literature, see e.g. Andersen and Andreasen (2000), Briani, Natalini, and Russo (2004), Cont and Tankov (2004), and d'Halluin, Forsyth, and Vetzal (2005). Although the methods are quite diverse, they all treat the integral and diffusion terms of the PIDE separately. Invariably, the integral term is evaluated explicitly in order to avoid solving a dense system of linear equations. In addition, the Fast Fourier Transform (FFT) algorithm is employed to speed up the computation of the integral term (which can be regarded as a convolution) and/or its inverse. Unfortunately, these methods require several approximations such as:

- in infinite activity processes, small jumps are approximated by a diffusion and incorporated into the diffusion term;

- the integral term must be localized to the bounded domain of the diffusion term, i.e. large jumps are truncated;

- the option price behaviour outside the solution domain must be assumed;

- the separate treatment of diffusion and integral components requires that function values are interpolated and extrapolated between the diffusion and integral grids in order to compute the convolution term.

These factors together make finite difference methods for option pricing under jump models quite complex, and potentially prone to accuracy and stability problems, especially for path dependent claims.

In section 3, we present a new Fourier Space Time-stepping (FST) algorithm. This method avoids the problems associated with finite difference methods by transforming the PIDE into Fourier space. One of the advantages of working directly in Fourier space is that the characteristic exponent of a independent increment stochastic process can be factored out of the Fourier transform of the PIDE. Consequently, the Fourier transform can be applied to the PIDE to obtain a linear system of easily solvable ordinary differential equations (ODE). Furthermore, the characteristic exponent is available, through the Lévy-Khintchine formula, in closed form for all independent increment processes. This makes the FST method quite flexible and generic – contingent claims on any exponential-Lévy stock price processes can then be priced with no additional modifications to the algorithm. The FST naturally leads to a symmetric treatment of the diffusion and jump terms and avoids any explicit assumptions on the option price outside of a truncated domain.

Since the FST method provides exact pricing results between monitoring times, it is significantly more efficient and accurate when compared with finite-difference methods for valuing Bermudan options. Furthermore, the method allows prices from one monitoring time to be projected back to a second monitoring time in one step of the algorithm. Contrastingly, finite-difference schemes will require time-stepping between monitoring dates resulting in further pricing biases and speed reduction.

For path independent options, prices for a range of spots can be obtained in a single time step. For exotic, path dependent options, we demonstrate how the FST method can handle barrier and Bermudan (American) styled clauses. The closed form expression for the Fourier transform of the option payoff is not required, making the FST method easily applicable to options with non-standard payoffs. Since the FST method requires two FFTs per time step, its computational complexity is $O(K\,N \log N)$, where $N$ is the number of spatial gridpoints and $K$ is the number of time steps. Through numerical experiments, we establish that the order of convergence of the method is two in space and one in time for path dependent options.

In section 4, we generalize the method to the multi-asset case and apply it to the specific example of spread options and comment on its use in catastrophe option pricing. In typical markets, jump models alone cannot match the implied volatility skew for longer maturities; however, the observed market behaviour can be captured by incorporating regime switches. This motivates us to include one more generalization and we introduce a non-stationary extension of the multi-dimensional Lévy processes using regime changes. The regime changes are induced through a homogenous continuous time Markov chain. This allows the index(es) to exhibit stochastic volatility and/or stochastic correlation behaviour which can be important in longer term options. Stochastic correlation has received little attention in the literature to date; however our modeling and pricing framework easily handles this feature.

We conclude this paper by discussing the possible applications, improvements and extensions of the FST algorithm.

## 2. Option Pricing with Differential Equations

In this section, we review the differential equation approach to option pricing with exponential Lévy processes. For a modern treatment of this subject the interested reader is referred to the monograph by Cont and Tankov (2004) and to Sato (1999) for further mathematical background.

Let $V(t, S(t))$ denote the price at time $t$ of an option, written on an underlying price index $S(t)$, with a $T$-maturity payoff of $\varphi(S(T))$. It is well known that, in an arbitrage-free and frictionless market, the value of the option is the discounted expectation under a, not necessarily unique, risk-neutral measure $\mathbb{Q}$ (see Harrison and Pliska (1981)). Explicitly,

$$V(t, S(t)) = \mathbb{E}_t^{\mathbb{Q}} \left[ e^{-r(T-t)} \varphi(S(T)) \right] , \tag{1}$$

where the expectation is taken with respect to the information, or filtration, $\mathcal{F}_t$ available at time $t$. Here and in the remainder of this article, we assume that the risk-free interest rate $r$ is constant. When the underlying index follows a diffusion process, the risk-neutral measure is indeed unique; however, in the more interesting case of exponential Lévy models, many equivalent risk-neutral measures exist. Nonetheless, we take the view that a trader is using such a model to price derivative instruments and therefore is modeling directly under a particular risk-neutral measure – possibly induced through a calibration procedure.

A dual and equivalent specification of the value function is its associated PIDE formulation. These two specifications are connected by noting that the discount-adjusted and log-transformed price process $v(t, X(t)) := e^{r(T-t)} V(t, S(0) e^{X(t)})$ is a martingale under the measure $\mathbb{Q}$. Consequently, the associated drift term of its defining SDE is identically zero. If the underlying index follows an exponential Lévy process, then the price process can be written as $S(t) = S(0) e^{X(t)}$ where $X(t)$ is a Lévy process with characteristic triplet $(\gamma, \sigma, \nu)$. In this case, the process $X(t)$ admits the following canonical Lévy-Itô decomposition into its diffusion and jump components

(see Sato (1999)):

$$X(t) = \gamma\,t + \sigma\,W(t) + J^l(t) + \lim_{\epsilon\searrow 0} J^\epsilon(t)\,, \tag{2}$$

$$J^l(t) = \int_0^t \int_{|y|\geq 1} y\,\mu(dy\times ds)\,, \tag{3}$$

$$J^\epsilon(t) = \int_0^t \int_{\epsilon\leq|y|<1} y\,[\mu(dy\times ds) - \nu(dy\times ds)]\,. \tag{4}$$

Here $W_t$ is a standard Brownian motion, $\mu(dy\times ds)$ is a Poisson random measure counting the number of jumps of size $y$ occurring at time $s$, and $\nu(dy\times ds) = \nu(dy)\,ds$ is its compensator. Note that $J^l(t)$ ($J^\epsilon(t)$) carries the interpretation of large (small) jumps. If the model has finite activity ($\int_{\mathbb{R}/\{0\}}(|y|\wedge 1)\,\nu(dy) < +\infty$), then there is no need to truncate small jumps and they can be lumped together with large jumps. We, however, choose to leave the decomposition general. Given this modeling assumption, $v$ must satisfy the PIDE

$$\begin{cases} (\partial_t + \mathcal{L})\,v &= 0\,, \\ v(T,x) &= \varphi(S(0)\,e^x)\,, \end{cases} \tag{5}$$

where $\mathcal{L}$ is the infinitesimal generator of the Lévy process and acts on twice-differentiable functions $f(x)$ as follows

$$\begin{aligned} \mathcal{L}f(x) &= \lim_{t\searrow 0} \frac{E[f(x+X(t))] - f(x)}{t} \\ &= \gamma\,\partial_x f + \tfrac{1}{2}\sigma^2\partial_{xx}f + \int_{\mathbb{R}/\{0\}} \left[f(x+y) - f(x) - y\,\mathbb{1}_{\{|y|<1\}}\,\partial_x f(x)\right]\nu(dy). \end{aligned} \tag{6}$$

By enforcing the risk-neutrality condition, the drift is uniquely determined once the volatility and Lévy density are specified. In particular, $\gamma$ satisfies

$$\mathbb{E}_0\left[e^{X(1)}\right] = e^r \qquad \Rightarrow \qquad \gamma = r - \Psi(-i)\,, \tag{7}$$

where $\Psi(\omega)$ denotes the characteristic exponent of the Lévy process and is provided explicitly by the Lévy-Khintchine formula (see Sato (1999))

$$\Psi(\omega) := \ln\mathbb{E}_t^{\mathbb{Q}}[e^{i\,\omega\,X(1)}] = i\,\gamma\,\omega - \tfrac{1}{2}\sigma^2\,\omega^2 + \int_{\mathbb{R}/\{0\}} (e^{i\omega y} - 1 - i\omega y\,\mathbb{1}_{\{|y|<1\}})\,\nu(dy). \tag{8}$$

Within this framework, the classical purely diffusive (BSM) model is recovered by setting the Lévy density to zero. Furthermore, jump-diffusion models, in which the log-stock price contains a diffusive component together with jumps occurring at Poisson times, are recovered by setting $\nu(dy) = \lambda\,f_Y(y)\,dy$ where $\lambda$ is the activity rate of the Poisson process and $f_Y(y)$ is the probability density of the jumps. In this case, the process $X(t)$ can be written in terms of a $\mathbb{Q}$-standard Brownian motion $W(t)$, a Poisson process $N(t)$ with activity rate $\lambda$, and i.i.d. random variables $Y_i$, representing the jumps at Poisson times $t_i$, as follows: $X(t) = \gamma\,t + \sigma\,W(t) + \sum_{n=1}^{N(t)} Y_n$. Two

| Model | Lévy density $\nu(dy)$ | Characteristic Exponent $\Psi(\omega)$ |
|---|---|---|
| BSM | N/A | $i(\mu - \frac{\sigma^2}{2})\omega - \frac{\sigma^2\omega^2}{2}$ |
| Merton JD | $\frac{\lambda}{\sqrt{2\pi\tilde{\sigma}^2}}e^{-\frac{1}{2}((y-\tilde{\mu})/\tilde{\sigma})^2}$ | $i(\mu - \frac{\sigma^2}{2})\omega - \frac{\sigma^2\omega^2}{2} + \lambda(e^{i\tilde{\mu}\omega - \tilde{\sigma}^2\omega^2/2} - 1)$ |
| Kou JD | $\lambda\left(p\,\eta_+\,e^{-y/\eta_+}\,\mathbb{1}_{y>0}\right.$ | |
| | $\left.+(1-p)\,\eta_-\,e^{-|y|/\eta_-}\,\mathbb{1}_{y<0}\right)$ | $i(\mu - \frac{\sigma^2}{2})\omega - \frac{\sigma^2\omega^2}{2} + i\omega\lambda(\frac{p}{\eta_+ - i\omega} - \frac{1-p}{\eta_- - i\omega})$ |
| VG | $\frac{1}{\kappa|y|}e^{\alpha\,y - \beta\,|y|}$ | $-\frac{1}{\kappa}\log(1 - i\mu\kappa\omega + \frac{\sigma^2\kappa\omega^2}{2})$ |
| NIG | $\frac{\gamma}{|y|}e^{\alpha\,y}\,K_1(\delta|y|)$ | $\frac{1}{\kappa} - \frac{1}{\kappa}\sqrt{1 - 2i\mu\kappa\omega + \sigma^2\kappa\omega^2}$ |
| CGMY | $\frac{C}{|x|^{1+Y}}\left(e^{-G|x|}\mathbb{1}_{x<0} + e^{-M\,x}\mathbb{1}_{x>0}\right)$ | $C\Gamma(-Y)\left[(M-i\omega)^Y - M^Y + (G+i\omega)^Y - G^Y\right]$ |

Table 1

The Lévy densities and characteristic exponent for various models. Here $\alpha = \frac{\mu}{\sigma^2}$, $\beta = \frac{\sqrt{\mu^2 + 2\sigma^2/\kappa}}{\sigma^2}$, $\gamma = \frac{\sqrt{\mu^2 + \sigma^2/\kappa}}{\pi\sigma\sqrt{\kappa}}$, $\delta = \frac{\sqrt{\mu^2 + \sigma^2/\kappa}}{\sigma^2}$ and $K_p(z)$ is the modified Bessel function of the second kind.

widely used jump-diffusion models are the log-normal jump model due to Merton (1976) and the double exponential model due to Kou (2002).

The Kou (2002) model assumes that the $X(t)$ jumps are double exponentially distributed, with positive jumps (with probability $p$) of mean size $\eta_+$ and negative jumps (with probability $1 - p$) of mean size $\eta_-$.

The characteristic exponents and Lévy densities for these models are provided in Table 1. These jump-diffusion models are popular not only because they perform well when calibrating to option prices, but also because they admit two semi-explicit closed form solutions. One form involves an infinite summation of BSM like prices (which can be safely truncated to a small number of terms), while the other form involves an inverse Fourier transformation. The interested reader is referred to the respective papers for details.

More recently, pure jump models have become very popular across a number of markets including equity, interest rate and commodity markets. These models have been found to better fit implied volatility smiles than jump-diffusion models and are widely used in industry. Huang and Wu (2004) carry out numerous statistical tests which demonstrate that models with infinitesimal jumps outperform jump-diffusion models. Within this class, the jumps themselves occur infinitely often with most jumps being of infinitesimal size. Several breeds of pure jump models have been suggested in the literature and each has its own merits and drawbacks. Three very popular models are the Variance-Gamma (VG) model of Madan and Seneta (1990) and Madan, Carr, and Chang (1998), the CGMY extension of the VG model developed by Carr, Geman, Madan, and Yor (2002), and the normal inverse Gaussian (NIG) model popularized by Barndorff-Nielsen (1997). The various Lévy densities and characteristic functions are provided in Table 1.

In the absence of jump components, the resulting PDE can be discretized using standard divided differences to approximate the first and second order derivatives. Whether the approximation scheme is carried out explicitly or implicitly or through a weighted scheme, the resulting system is tri-diagonal and leads to very efficient numerical approximations. Unfortunately, when jumps are present, the integral term in (5) must be approximated resulting in a dense matrix structure. Several methods for dealing with this issue have been presented in the literature with most relying on the explicit evaluation of approximations to the integral term in conjunction with an iterative refinement and possibly FFT speedup.

Andersen and Andreasen (2000) propose an FFT-Alternating Direction Implicit (FFT-ADI) method which treats the diffusion and integral terms symmetrically over a full time step by splitting the time step into two half-steps; an explicit scheme is used on the first half-step and implicit scheme on the second half-step. The inversion of the dense matrix is performed efficiently by regarding the term as a convolution and utilizing the FFT algorithm. The fixed point iteration scheme of d'Halluin, Forsyth, and Vetzal (2005) treats the integral term explicitly while iterating to attain required error tolerance per time-step. In addition, Implicit-Explicit (IMEX) Runge-Kutta schemes have been applied in Briani, Natalini, and Russo (2004) to solve the PIDEs. Although the FFT algorithm is frequently used to speed up the computation of the integral term, such schemes require careful mapping of function values between the diffusion and integral grids. We circumvent the problems posed by working in real space and instead opt to solve the problem directly in Fourier space as explained in the next section.

### 3. Fourier Space Time-stepping

Fourier and Laplace transforms have been used extensively to solve PDEs, either by transforming the equation into an ODE or expressing the solution as an infinite series (see Strauss (1992) and Taylor (1997)). The aim of this section is to develop a Fourier transform methodology for solving PIDEs of the form (5). The main advantage of transform methods is that the PIDE can be handled efficiently without the additional complexities associated with the integral term. Additionally, the algorithm is applicable to any independent increment stock price model which admits a closed form characteristic function. Furthermore, we extend this approach to the valuation of path dependent options, such as barrier, American, and shout options, and discuss the convergence of these numerical schemes.

### 3.1. Transforming the PIDE

A Pseudo Differential Operator (PDO) extends the notion of a differential operator and is widely used to solve differential equations. The essential idea is that a differential operator with constant coefficients can be represented as a composition of a Fourier transform, multiplication by a polynomial function, and an inverse Fourier transform. Only a few fundamental facts from PDO theory are required to derive our numerical method. The interested reader is referred to Boyarchenko and Levendorskii (2002) who discuss the PDO theory in the context of option pricing. For a more thorough treatment of the subject see Taylor (1997).

A function in the space domain $f(x)$ can be transformed to a function in the frequency domain $\mathcal{F}[f](\omega)$ (where $\omega$ is given in radians per second) and vice-versa using the Fourier transform:

$$\mathcal{F}[f](\omega) := \int_{-\infty}^{\infty} f(x) e^{-i\omega x} dx \quad \text{and} \quad \mathcal{F}^{-1}[\hat{f}](x) := \frac{1}{2\pi} \int_{-\infty}^{\infty} \widehat{f}(\omega) e^{i\omega x} d\omega.$$

The Fourier transform is a linear operator that maps spatial derivatives $\partial_x$ into multiplications in the frequency domain:

$$\mathcal{F}\left[\partial_x^n f\right](t, w) = iw\, \mathcal{F}\left[\partial_x^{n-1} f\right](t, w) = \cdots = (iw)^n \mathcal{F}[f](t, w).$$

Consequently, applying the Fourier transform to the infinitesimal generator $\mathcal{L}$ of $X(t)$, defined by equation (6), allows the characteristic exponent of $X(t)$ to be factored out:

$$
\begin{aligned}
\mathcal{F}[\mathcal{L}v](t, \omega) &= \left\{ i\gamma\omega - \frac{\sigma^2 \omega^2}{2} + \int_{\mathbb{R}/\{0\}} [e^{i\omega x} - 1 - i\omega y \mathbb{1}_{\{|y|<1\}}] \nu(dy) \right\} \mathcal{F}[v](t, \omega) \\
&= \Psi(\omega) \mathcal{F}[v](t, \omega).
\end{aligned}
\tag{9}
$$

Furthermore, taking the Fourier transform of both sides of the PIDE (5) leads to

$$
\begin{cases}
\partial_t \mathcal{F}[v](t, \omega) + \Psi(\omega)\mathcal{F}[v](t, \omega) &= 0\,, \\
\mathcal{F}[v](T, \omega) &= \mathcal{F}[\varphi](\omega)\,.
\end{cases}
\tag{10}
$$

The PIDE is therefore transformed into a one-parameter family of ODEs (10) parameterized by $\omega$. Giving the value at time $t_2 \leq T$, the system is easily solved to find the value at time $t_1 < t_2$:

$$\mathcal{F}[v](t_1, \omega) = \mathcal{F}[v](t_2, \omega) \cdot e^{(t_2 - t_1)\Psi(\omega)}.\tag{11}$$

Taking the inverse transform leads to the final result

$$v(t_1, x) = \mathcal{F}^{-1}\left\{ \mathcal{F}[v](t_2, \omega) \cdot e^{(t_2 - t_1)\Psi(\omega)} \right\}(x).\tag{12}$$

### 3.2. Direct Transform Method

Alternatively, it is possible to derive (12) directly from the expectation representation of prices. Recall that $v$ is a $\mathbb{Q}$ martingale; consequently,

$$
\begin{aligned}
v(t_1, X(t_1)) &= \mathbb{E}_{t_1}^{\mathbb{Q}}\left[v(t_2, X(t_2))\right] \\
&= \int_{-\infty}^{\infty} v(t_2, X(t_1) + x)\, f_{X(t_2) - X(t_1)}(x)\, dx \\
&= \int_{-\infty}^{\infty} v(t_2, X(t_1) + x)\, f_{X(t_2 - t_1)}(x)\, dx\,.
\end{aligned}
$$

Here, $f_{X(t)}(x)$ denotes the p.d.f. of the process $X(t)$ and the third line follows from the independent increment property of the process $X(t)$. Furthermore, $\mathcal{F}[f_{X(t)}](\omega) = e^{t\,\Psi(-\omega)}$ and since a convolution in real space corresponds to multiplication in Fourier space, we have

$$\mathcal{F}[v](t_1, \omega) = \mathcal{F}[v](t_2, \omega) \, e^{(t_2 - t_1) \, \Psi(\omega)} \, . \tag{13}$$

### 3.3. FST Method

Armed with the pseudo-differential operator solution (12), the numerical algorithm is straightforward. For path-independent options the price is obtained in one step by directly applying equation (12) similar in spirit to Carr and Madan (1999). For path dependent options a time-stepping algorithm is used to apply boundary conditions, impose constraints, or optimize over a policy domain.

Consider a partition of the time and truncated stock price domain $\Omega = [0, T] \times [x_{min}, x_{max}]$ into a finite mesh of points $\{t_n | n = 0, \ldots, N\} \times \{x_m | m = 0, \ldots, M-1\}$, where $t_n = n\Delta t$, $x_m = x_{min} + m\Delta x$ and $\Delta t = T/N$, $\Delta x = (x_{max} - x_{min})/M$. The mesh boundary $\{x_{min}, x_{max}\}$ is chosen large enough to capture the overall behaviour of the option value function, yet small enough to maintain the accuracy of the computed option price in the range of interest. Numerical experiments suggest that $x_{min} \in [-4, -2]$ and $x_{max} \in [2, 4]$ works well for diffusion models, while $x_{min} \in [-5, -3]$ and $x_{max} \in [3, 5]$ is preferable for models with a dominant jump component. Note that $x = \log(S/S_0)$. Analogously, consider a partitioning of the time and frequency domain $\widehat{\Omega} = [0, T] \times [0, \omega_{max}]$ into a finite mesh of points $\{t_n | n = 0, \ldots, N\} \times \{\omega_m | m = 0, \ldots, M/2\}$, where $\omega_m = m\Delta\omega$ and $\Delta\omega = 2\omega_{max}/M$. We choose $\omega_{max} = \frac{1}{2\Delta x}$, which is the Nyquist critical frequency. Note that $v(t, x)$ is a real-valued function and thus $\mathcal{F}[v](t, -\omega) = \overline{\mathcal{F}[v](t, \omega)}$. The Fourier transform for negative frequencies is not computed and therefore the frequency grid has half as many points as the spatial grid.

Let $v_m^n := v(t_n, x_m)$ represent $v(t, x)$ at the node points of the partition of $\Omega$, and let $\hat{v}_m^n := \hat{v}(t_n, \omega_m)$ represent $\mathcal{F}[v](t, \omega)$ at the node points of the partition $\widehat{\Omega}$. The frequency domain prices are obtained from the spatial domain prices as follows:

$$\begin{aligned}
\hat{v}_m^n = \mathcal{F}[v](t_n, \omega_m) &\approx \sum_{k=0}^{M-1} v(t_n, x_k) e^{-i\omega_m x_k} \Delta x \\
&= \alpha_m \sum_{k=0}^{M-1} v_k^n e^{-imk/M} \\
&= \alpha_m \mathsf{FFT}[v^n](m).
\end{aligned} \tag{14}$$

Here, $\alpha_m = e^{-i\omega_m x_{min}} \Delta x$ and $\mathsf{FFT}[v^n](m)$ denotes the $m$-th component of the discrete Fourier transform (DFT) of the vector $v^n$, which can be computed efficiently using the FFT algorithm. Similarly, the spatial domain prices can be computed from frequency domain prices via a discrete inverse transform

$$v_m^n = \mathsf{FFT}^{-1}[\alpha^{-1} \cdot \hat{v}^n](m). \tag{15}$$

Combining these connections between frequency and spatial domains with the transformed

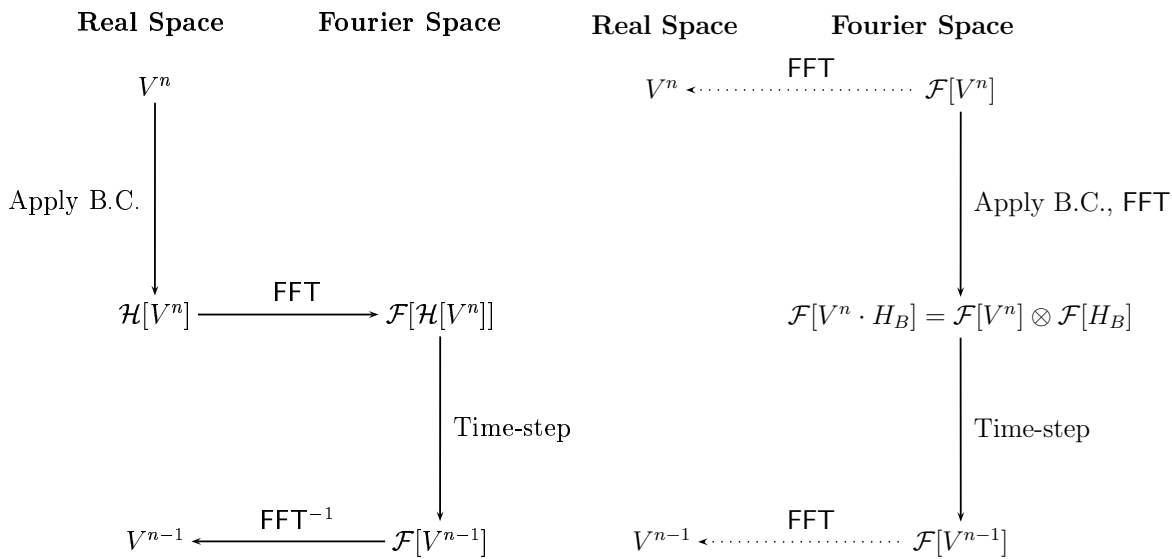

Figure 1. A schematic representation of the FST algorithm. In the left panel, the boundary conditions (such as optimal exercise or barrier breach) are applied in real space while the time step is performed in Fourier space. In the right panel, the refined algorithm for Barrier breach permits application of boundary conditions in Fourier space for particular cases.

PIDE (12), a step backwards in time is computed by

$$
\begin{aligned}
v^{n-1} &= \mathsf{FFT}^{-1}[\alpha^{-1} \cdot \hat{v}^{n-1}] \\
&= \mathsf{FFT}^{-1}[\alpha^{-1} \cdot \hat{v}^{n} \cdot e^{\Psi \, \Delta t}] \\
&= \mathsf{FFT}^{-1}[\alpha^{-1} \cdot \alpha \cdot \mathsf{FFT}[v^{n}] \cdot e^{\Psi \, \Delta t}] \\
&= \mathsf{FFT}^{-1}[\mathsf{FFT}[v^{n}] \cdot e^{\Psi \, \Delta t}].
\end{aligned}
\tag{16}
$$

Notice that the coefficient $\alpha$, which embeds information about the spatial boundary, cancels in the above equation and can be omitted during the numerical computation. Certain payoffs contain singularities in their Fourier transforms along the real axis. A simple shifting of $\omega \to \omega + i\epsilon$ avoids this problem, resulting in a slight modification of the time-stepping algorithm: $v^{n-1} = \mathsf{FFT}^{-1}[\mathsf{FFT}[\check{v}^{n}] \cdot e^{\check{\Psi} \, \Delta t}]$ where $\check{v}^{n}_{k} = e^{\epsilon x_{k}} v^{n}_{k}$ and $\check{\Psi}(\omega) = \Psi(\omega + i\epsilon)$.

### 3.3.1. European Options

European options can be valued in a single time step, since (16) is a valid approximation for any $\Delta t$. In this case, given a payoff function $\varphi(S)$, set $N = 1$, $v^{1}_{m} = \varphi(S(0)e^{x_{m}})$, numerically invert $v^{1}_{m}$ to obtain $\hat{v}^{1}_{m}$ via (14), and finally apply (16). This approach is similar to Carr and Madan (1999), however, the explicit expression of the Fourier transformed option payoff is not required – clearly a great advantage for non-standard payoffs. Moreover, our approach is computationally more efficient when compared to spatial PIDE solution-based methods since it does not require stepping in time.

### 3.3.2. American Options

American options can be priced using a finite difference method either by solving a linear complementarity problem (see Dewynne, Howison, and Wilmott (1993), Dempster and Hutton (1997), Huang and Pang (1998), and Forsyth and Vetzal (2002)), or solving a free boundary value problem (see McKean (1965), Kim (1990), and Carr, Jarrow, and Myneni (1992)). Although the free boundary formulation for American options is an active area of research and can be potentially combined with the FST method, the linear complementarity formulation is easier to implement in the context of the FST method. Since the value of an American option is always greater than or equal to the terminal payoff, the idea is to continuously enforce the condition $V(t, S) \geq V(T, S)$. Numerically, this is enforced when boundary conditions are applied, resulting in the following algorithm:

$$v^{n-1} = \max \left\{ \mathsf{FFT}^{-1}[\mathsf{FFT}[v^n] \cdot e^{\Psi \Delta t}], v^N \right\}. \tag{17}$$

where the $\max(\cdot, \cdot)$ is taken componentwise. There is no convenient representation of the $\max(\cdot, \cdot)$ operator in Fourier space; consequently, it is necessary to switch between real and Fourier spaces at each time-step. Schematically, the algorithm is presented in the left panel of Figure 1.

### 3.3.3. Barrier Options

The numerical algorithm for barrier options is similar to that of American options and also involves enforcement of constraints. Here, we discuss the up-and-out barrier option case; however, the results can be extended to other barrier option styles. In spatial coordinates, the barrier boundary condition forces

$$V(t, S) = R \quad \text{for } S \geq B, \tag{18}$$

where $B$ is the knock-out barrier level and $R$ is the rebate paid in the case of knock-out. In terms of the time stepping algorithm:

$$v^{n-1} = \mathsf{FFT}^{-1}[\mathsf{FFT}[v^n] \cdot e^{\Psi \Delta t}] \cdot H_B, \tag{19}$$

where $H_B(x) = \mathbb{1}_{\{x < \ln(B/S(0))\}} + R \cdot \mathbb{1}_{\{x \geq \ln(B/S(0))\}}$. By noting that the Fourier transform of a product of two functions is the convolution of their respective Fourier transforms, it is possible to lift the algorithm to the frequency domain entirely. Thus, the time-stepping algorithm can be modified to

$$\hat{v}^{n-1} = \{\mathsf{FFT}[H_B]\} \otimes \{\hat{v}^n \cdot e^{\Psi \Delta t}\}. \tag{20}$$

where the convolution of two vectors, denoted by $\otimes$, can be executed efficiently using the FFT algorithm. This approach requires only a single FFT operation (since $\mathsf{FFT}[H_B]$ is precomputed) per time-step as compared to two operations in the original approach. This is schematically represented in the right panel of Figure 1.

Numerical experiments show that direct application of (19) results in erratic convergence, especially near the barrier. To stabilize the algorithm we enforce constraint (18) numerically via the method of images (see e.g. Buchen (1996)) by truncating the values of $V(t, S)$ at $S = B$ and extending it to an odd function, i.e. setting $V(t, B + x) = 2R - V(t, B - x)$ for $x > 0$ and $V(t, B) = R$. This procedure is repeated at each time step and, without introducing any bias into the solution of the equation for $S \leq B$, it stabilizes the convergence of the FST algorithm (see Table 15).

### 3.3.4. Shout Options

Shout options constitute a significantly more complicated but interesting class of problems for numerical valuation. Much like American options, the holder of a shout option may exercise their right at any time during the contract lifetime. Specifically, the contract allows the holder to "shout" and set a new strike level. The shout regions are obtained by solving a dynamic programming problem at each time step similar to computing the optimal exercise boundary of American options. An $n$-shout option allows the investor to "shout" and reset the strike price $n$ times, requiring a more subtle analysis in pricing of such options. For $n$-shout options, the FST algorithm must be extended to track several prices at once (each price corresponding to the number of shouts remaining). Details of the application of the FST algorithm to these options is forthcoming in Jackson, Jaimungal, and Surkov (2007).

### 3.4. Numerical Results

Below we present our pricing results and compare them to the prices found in the literature. The option and stock price models are specified below each table. The closed-form price refers to the option price calculated by an analytic formula, if available, while quoted price refers to the price calculated by the authors of the paper. Note that the quoted result is the most precise value given in the source and not the value that the algorithm converges to. If a closed-form formula is not available, a very good approximation for European options can be found by evaluating the integral form in Carr and Madan (1999) using an adaptive quadrature method. This price is referred to as the integral price.

It is also of great interest to establish the convergence properties of the FST algorithm. Here we follow the estimation approach taken by d'Halluin, Forsyth, and Vetzal (2005) and extend it to estimate the order as a function of $\Delta t$ and $\Delta x$ independently. First, assume that

$$v_{approx}(\Delta t, \Delta x) = v_{exact} + c_t(\Delta t)^{p_t} + c_x(\Delta x)^{p_x},$$

where $c_t$, $p_t$, $c_x$, and $p_x$ are constants. Since the algorithm does not require time-stepping to value European options, the equation above can be simplified to depend only on $\Delta x$,

$$v_{approx}(\Delta x) = v_{exact} + c_x(\Delta x)^{p_x}.$$

| N | Value | Change | $\log_2$Ratio | CPU-Time |
|---|-------|--------|--------------|----------|
| 2048 | 18.00329705 | | | 0.002712 |
| 4096 | 18.00354600 | 0.000249 | | 0.003550 |
| 8192 | 18.00360820 | 0.000062 | 2.0008 | 0.007105 |
| 16384 | 18.00362375 | 0.000016 | 2.0004 | 0.014658 |
| 32768 | 18.00362764 | 0.000004 | 2.0002 | 0.026558 |

Table 2

*Option:* European put $S = 100.0, K = 100.0, T = 10$; *Model:* Merton jump-diffusion $\sigma = 0.15, \lambda = 0.1, \tilde{\mu} = -1.08, \tilde{\sigma} = 0.4, r = 0.05, q = 0.02$; *Closed-Form Price:* 18.003629 *Quoted Price:* 18.0034 *Source:* Andersen and Andreasen (2000)

We use estimates of the option price $v_{approx}$ on successively finer grids in space to establish the rate of convergence via

$$p_x = \log_2 \frac{|v_{approx}(\Delta x) - v_{approx}(\Delta x/2)|}{|v_{approx}(\Delta x/2) - v_{approx}(\Delta x/4)|}. \tag{21}$$

Here, the absolute changes in the numerator and the denominator are given in the table under the column "Change", while the estimated rate of convergence is given under the column "$\log_2$Ratio".

For European options under various processes (see Table 2 and Tables 7 - 9 in Appendix A) we find that the FST algorithm is order 2 in the space variable. For path dependent options it is also necessary to establish convergence properties of the algorithm in the time variable. By holding $\Delta x$ constant, the error becomes dependent only on $\Delta t$. We assume

$$v_{approx}(\Delta t) = v_{exact} + c_t(\Delta t)^{p_t}$$

with $p_t$ estimated by

$$p_t = \log_2 \frac{|v_{approx}(\Delta t) - v_{approx}(\Delta t/2)|}{|v_{approx}(\Delta t/2) - v_{approx}(\Delta t/4)|}. \tag{22}$$

Results of estimating $p_t$ are presented in Appendix B and overwhelmingly suggest that the FST algorithm is order 1 in the time dimension. Estimating $p_t$ and $p_x$ independently leads us to believe that a good estimate for the error in the FST algorithm is

$$v_{approx}(\Delta t, \Delta x) = v_{exact} + c(\Delta)^p,$$

where $\Delta t = O(\Delta^2)$ and $\Delta x = O(\Delta)$. Again $p$ can be estimated by computing the log ratio of changes in $v_{approx}$ as $\Delta t$ is reduced by the square of the relative reduction in $\Delta x$,

$$p = \log_2 \frac{|v_{approx}(\Delta t, \Delta x) - v_{approx}(\Delta t/4, \Delta x/2)|}{|v_{approx}(\Delta t/4, \Delta x/2) - v_{approx}(\Delta t/16, \Delta x/4)|}. \tag{23}$$

| N | M | Value | Change | $\log_2$Ratio | CPU-Time |
|---|---|-------|--------|------------|----------|
| 2048 | 128 | 9.22186786 | | | 0.031823 |
| 4096 | 512 | 9.22447378 | 0.002606 | | 0.169965 |
| 8192 | 2048 | 9.22520088 | 0.000727 | 1.8416 | 1.756965 |
| 16384 | 8192 | 9.22538212 | 0.000181 | 2.0042 | 13.571508 |
| 32768 | 32768 | 9.22542570 | 0.000044 | 2.0564 | 156.621382 |

Table 3

*Option:* American put $S = 90.0, K = 98.0, T = 0.25$; *Model:* CGMY $C = 0.42, G = 4.37, M = 191.2, Y = 1.0102, r = 0.1$; *Quoted Price:* 9.2185 *Source:* Forsyth, Wan, and Wang (2006)

| N | M | Value | Change | $\log_2$Ratio | CPU-Time |
|---|---|-------|--------|------------|----------|
| 2048 | 128 | 0.25348926 | | | 0.016819 |
| 4096 | 512 | 0.25401054 | 0.000521 | | 0.181466 |
| 8192 | 2048 | 0.25414993 | 0.000139 | 1.9030 | 1.297593 |
| 16384 | 8192 | 0.25418356 | 0.000034 | 2.0514 | 10.794937 |
| 32768 | 32768 | 0.25419310 | 0.000010 | 1.8178 | 123.635629 |

Table 4

*Option:* Up-and-Out Barrier Call $S = 100.0, K = 100.0, B = 110, T = 1.0$; *Model:* Black-Scholes-Merton $\sigma = 0.15, r = 0.05, q = 0.02$; *Closed-Form Price:* 0.2541963 *Source:* Hull (2005)

| N | M | Value | Change | $\log_2$Ratio | CPU-Time |
|---|---|-------|--------|------------|----------|
| 2048 | 128 | 35.73104311 | | | 0.719093 |
| 4096 | 512 | 35.88365480 | 0.152612 | | 4.285660 |
| 8192 | 2048 | 35.92030459 | 0.036650 | 2.0580 | 31.988542 |
| 16384 | 8192 | 35.93057212 | 0.010268 | 1.8357 | 286.700219 |

Table 5

*Option:* Shout $S = 100.0, T = 10.0, n = 5$; *Model:* CGMY $C = 1.0, G = 5.0, M = 5.0, Y = 0.5, r = 0.1$.

The results presented in Tables 3 - 5 and in Tables 10 - 12 in Appendix A support the conclusion that the error in the FST algorithm behaves as follows

$$v_{approx}(\Delta t, \Delta x) = v_{exact} + c_t \Delta t + c_x (\Delta x)^2. \qquad (24)$$

Our implementation of the FST algorithm uses the FFTW library developed by Frigo and Johnson (2005) to compute the FFTs. The numerical experiments were written in C++ and performed on a Pentium 4 1.50 GHz machine.

## 4. Extension to Multi-Asset Problems

Pricing of contingent claims that depend on two or more assets is a complicated task. Typical approaches include Monte-Carlo simulations, solving multi-dimensional PDEs and FFT based approaches. Although very flexible, Monte-Carlo methods can suffer from slow convergence when jumps are incorporated into the model. Solving a multi-dimensional PDE is more efficient; however, when jumps are introduced, the resulting PIDEs are difficult to handle due to the non-local integral term. Based on the FFT approach of Carr and Madan (1999), Dempster and Hong (2000) develop a method for pricing spread options – options on two underlying stocks. Unfortunately, their method is only applicable to European options and does not readily extend to path-dependent options.

The computational efficiency of the FFT is not restricted to a single dimension; as such, the multi-dimensional FFT can be leveraged to efficiently compute multi-dimensional DFT. Therefore, it seems natural to generalize our method to a multi-dimensional FST algorithm. The advantages of our approach over previous approaches is its ease in handling any Lévy process and its ease of application to path-dependent options. In this section, we present the main ideas of the derivation and conclude with pricing results for two-asset European, American and barrier options. The mathematical basis of our method follows through from Section 2.

Let $V(t, \mathbf{S}(t))$ denote the price at time t of an option, written on a vector of $d$ underlying price indices $\mathbf{S}(t)$, whose components are $S_j(t)$, with a $T$-maturity payoff of $\varphi(\mathbf{S}(T))$. The value of the option is the discounted expectation under the risk-neutral measure $\mathbb{Q}$

$$V(t, \mathbf{S}(t)) = \mathbb{E}_t^{\mathbb{Q}} \left[ e^{-r(T-t)} \varphi(\mathbf{S}(T)) \right] \ . \tag{25}$$

The discount-adjusted and log-transformed price process $v(t, \mathbf{X}(t)) := e^{r(T-t)} V(t, \mathbf{S}(0) e^{\mathbf{X}(t)})$ is a $\mathbb{Q}$-martingale. We assume that the underlying indices $\mathbf{S}(t)$ follow a $d$-dimensional exponential Lévy process with a characteristic triplet $(\boldsymbol{\gamma}, \mathbf{C}, \boldsymbol{\nu})$, where $\boldsymbol{\gamma}$ represents the vector of unadjusted-drifts, $\mathbf{C}$ represents the variance-covariance matrix of the diffusions, and $\boldsymbol{\nu}$ is the multi-dimensional Lévy density. As before, $\mathbf{X}$ admits a canonical Lévy-Itô decomposition

$$\mathbf{X}(t) = \boldsymbol{\gamma} \, t + \mathbf{W}(t) + \mathbf{J}^l(t) + \lim_{\epsilon \searrow 0} \mathbf{J}^\epsilon(t) \ , \tag{26}$$

$$\mathbf{J}^l(t) = \int_0^t \int_{|y| \geq 1} \mathbf{y} \, \mu(d\mathbf{y} \times ds) \ , \tag{27}$$

$$\mathbf{J}^\epsilon(t) = \int_0^t \int_{\epsilon \leq |y| < 1} \mathbf{y} \, [\mu(d\mathbf{y} \times ds) - \nu(d\mathbf{y} \times ds)] \ . \tag{28}$$

Applying the zero-drift condition on $v$ together with its boundary condition at maturity leads to the PIDE

$$\begin{cases} (\partial_t + \mathcal{L}) \, v = 0 \ , \\ v(T, \mathbf{x}) = \varphi(\mathbf{S}(0) \, e^{\mathbf{x}}) \ , \end{cases} \tag{29}$$

where $\mathcal{L}$ is the infinitesimal generator of the multi-dimensional Lévy process and acts on twice differentiable functions $f(\mathbf{x})$ as follows

$$\mathcal{L}f(\mathbf{x}) = \left(\boldsymbol{\gamma} \cdot \partial_{\mathbf{x}} + \tfrac{1}{2}\,\partial_{\mathbf{x}} \cdot \mathbf{C} \cdot \partial_{\mathbf{x}}\right) f(\mathbf{x}) + \int_{\mathbb{R}^n/\{\mathbf{0}\}} \left(f(\mathbf{x}+\mathbf{y}) - f(\mathbf{x}) - \mathbf{y} \cdot \partial_{\mathbf{x}} f(\mathbf{x})\, \mathbb{1}_{|\mathbf{y}|<1}\right) \nu(d\mathbf{y}) \quad (30)$$

A $d$-parameter family of ODEs is obtained by applying Fourier transforms to both sides of (29)

$$\partial_t \mathcal{F}[v](t,\boldsymbol{\omega}) + \Psi(\boldsymbol{\omega}) \mathcal{F}[v](t,\boldsymbol{\omega}) = 0 \;. \tag{31}$$

Here,

$$\Psi(\boldsymbol{\omega}) = \boldsymbol{\gamma} \cdot \boldsymbol{\omega} + \tfrac{1}{2}\,\boldsymbol{\omega} \cdot \mathbf{C} \cdot \boldsymbol{\omega} + \int_{\mathbb{R}^n} \left(e^{i\boldsymbol{\omega}\cdot\mathbf{y}} - 1 - i\mathbf{y}\cdot\boldsymbol{\omega}\,\mathbb{1}_{|\mathbf{y}|<1}\right) \nu(d\mathbf{y}) \tag{32}$$

is the characteristic exponent of the $d$-dimensional Lévy process. The system of ODEs admits the simple analytic solution

$$\mathcal{F}[v](t_1,\boldsymbol{\omega}) = \mathcal{F}[v](t_2,\boldsymbol{\omega})\, e^{(t_2-t_1)\Psi(\boldsymbol{\omega})} \tag{33}$$

resulting in a natural generalization of the FST time-stepping method to arbitrary dimensions

$$v^{n-1} = \mathsf{FFT}^{-1}[\mathsf{FFT}[v^n] \cdot e^{\Psi\,\Delta t}] \;. \tag{34}$$

Of course, this algorithm could be derived without going through the PIDE, and instead relying solely on the martingale property of $v$.

If jumps in individual assets are uncorrelated, then the multi-dimensional Lévy density factors $\boldsymbol{\nu}(d\mathbf{y} \times ds) = \nu_1(dy_1)\ldots\nu_n(dy_n)\,ds$ , leading to a factorization of the characteristic function. However, allowing for correlation between the jumps in the assets is just as straightforward as long as the integral appearing in (32) can be computed analytically in closed form. In an extreme case, there may be a single market wide jump risk factor together with idiosyncratic jump risks. In the next subsection we illustrate two applications of the FST algorithm in two-dimensions.

### 4.1. Spread Options

An interesting class of multi-asset options are spread options – the option to exchange $\beta$-units of one asset for $\alpha$-units of another asset. These options can be viewed as options on the difference (or spread) of two stock prices with terminal payoff

$$\varphi(S_1(T), S_2(T)) = \max(\alpha S_2(T) - \beta S_1(T) - K, 0) \;. \tag{35}$$

Spread options do not admit an analytic closed-form solution even for the Black-Scholes-Merton model. For a detailed discussion of spread options and various approximations see Carmona and Durrleman (2003). Dempster and Hong (2000) present an FFT-based approach to valuation of spread options. Their approach involves breaking the region in which the option is in-the-money into a series of rectangular approximations. Unfortunately, they only apply their method to a pure diffusion model with stochastic volatility and it seems difficult to extend this method to the Bermudan or barrier cases.

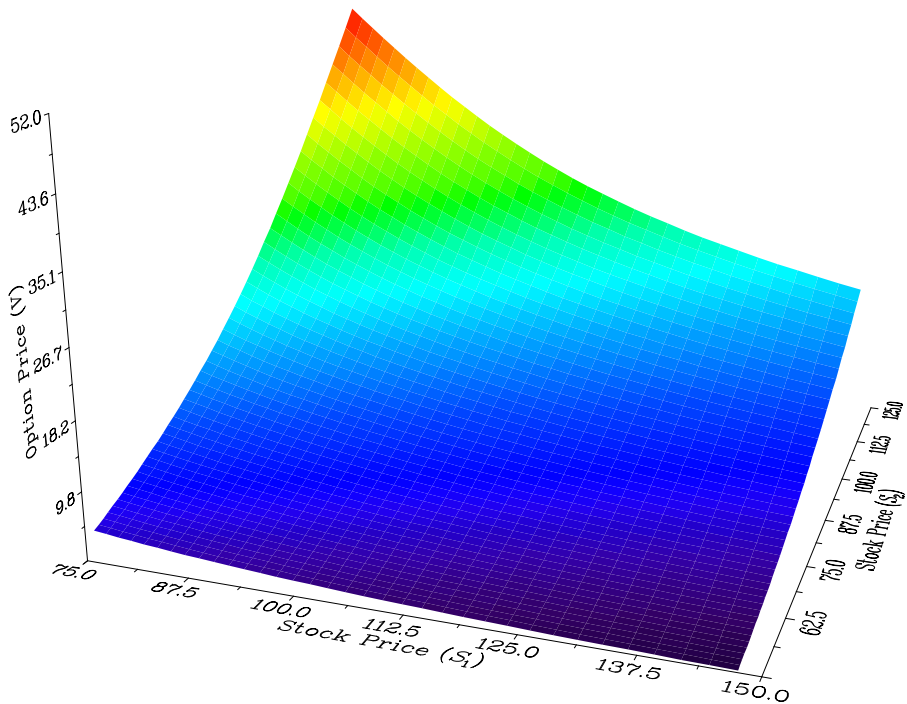

Figure 2. Price surface of a European spread call with parameters $T = 1, \alpha = 1.0, S_1 = 96, \beta = 1.0, S_2 = 100, K = 0$. The stock price process is Merton jump-diffusion $r = 0.1, \sigma_1 = 0.1, q_1 = 0.05, \lambda_1 = 0.25, \tilde{\mu}_1 = -0.13, \tilde{\sigma}_1 = 0.37, \sigma_2 = 0.2, q_2 = 0.05, \lambda_2 = 0.5, \tilde{\mu}_2 = 0.29, \tilde{\sigma}_2 = 0.41, \rho = 0.5$

For our numerical experiments, we assume a joint jump-diffusion with Merton-like jumps and compare our results with the Kirk (1995) approximation and its extension for jump-diffusions found in Carmona and Durrleman (2003). In this case, the Lévy density factors with $\nu_i(dy) = (\lambda_i/\sqrt{2\pi\tilde{\sigma}_i^2}) \exp\{-(y-\tilde{\mu}_i)^2/2\tilde{\sigma}_i^2\}dy$ and the diffusive volatilities are $\sigma_i$ with correlation $\rho$. Hence,

$$
\begin{aligned}
\Psi(\omega_1, \omega_2) &= i(\mu_1 - \frac{\sigma_1^2}{2})\omega_1 + i(\mu_2 - \frac{\sigma_2^2}{2})\omega_2 - \frac{\sigma_1^2\omega_1^2}{2} - \rho\sigma_1\sigma_2\omega_1\omega_2 - \frac{\sigma_2^2\omega_2^2}{2} \\
&\quad + \lambda_1\left(e^{i\tilde{\mu}_1\omega_1 - \tilde{\sigma}_1^2\omega_2^2/2} - 1\right) + \lambda_2\left(e^{i\tilde{\mu}_2\omega_2 - \tilde{\sigma}_2^2\omega_2^2/2} - 1\right)
\end{aligned}
\tag{36}
$$

where the drift is fixed by risk-neutrality to be $\mu_i = r - \lambda_i(e^{\tilde{\mu}_i + \tilde{\sigma}_i^2/2} - 1)$. In Appendix C we present numerical results for European and American options. Of course, the method is applicable to barrier spread options as well.

### 4.2. Catastrophe Options

Catastrophe options have become more important in recent years, yet there is very little published on the subject. These options pay the holder a function of total losses and the company's equity value. As a result, it is important to jointly model losses and equity, especially since large losses will cause significant drops in share value. Cox, Fairchild, and Pedersen (2004) introduce a simple model which counts the number of losses paying no attention to loss size. Jaimungal and Wang (2006) extend the model to incorporate random losses as well as stochastic interest rates. However, both works assume that the option is European, while the existing contracts are in fact of American type. Here, we illustrate how the FST can be used to price the early exercise premium. In the constant interest case, Jaimungal and Wang (2006) make the following assumption on share value $S(t)$ and losses $L(t)$

$$S(t) = S(0) \exp \{ J(t) + \gamma t + \sigma W_t \} , \tag{37}$$

$$J(t) = -\alpha L(t) , \tag{38}$$

$$L(t) = \sum_{n=1}^{N(t)} l_i , \tag{39}$$

where $N(t)$ is a compound Poisson process with activity rate $\lambda$, $l_i$ are i.i.d. random variables with probability density $f_L(l)$ having support on $\mathbb{R}_+$. Notice that it is the presence of losses which drives the jumps in the price process and not an independent jump process.

In this case, the 2-dimensional Lévy density is $\nu(dy_1 \times dy_2) = f_L(y_2) \, \delta(y_1 + \alpha \, y_2) \, dy_1 dy_2$ resulting in the characteristic function

$$\Psi(\omega_1, \omega_2) = i \, \gamma \, \omega_1 - \tfrac{1}{2} \sigma^2 \, \omega_1^2 + \int_{-\infty}^{\infty} \left( e^{i(-\alpha \, \omega_1 + \omega_2) \, y} - 1 - i \, y \, (-\alpha \omega_1 + \omega_2) \, \mathbb{1}_{|y| < 1} \right) dy . \tag{40}$$

The risk-neutral drift is $\gamma = r - \Psi(-i, 0)$. With this characteristic function a slight modification of the American styled options algorithm leads to an efficient pricing mechanism. Namely, the exercise policy is now chosen at each point in the $(S(t), L(t))$ plane independently. If $L(t)$ was not a separate observable, as it is in the usual jump-diffusion model case, then the exercise policy would be independent of $L(t)$. We leave the application of the FST method to catastrophe options such as catastrophe equity put options (with payoff $\varphi(S(T), L_T) = \mathbb{1}_{L_T > L_t + U} (K - S(T))_+$) for future work.

## 5. Regime Switching Models

Regime switching models can be traced back to the early work of Lindgren (1978) and ever since the seminal work of Hamilton (1989, 1990) have become a very popular approach to incorporate non-stationary behaviour into an otherwise stationary model. The essential idea is to assume that the world switches between states representing, for example, moderate, low and high volatilities regimes. Although popular for describing time-series, little work has been carried out in terms of option valuation. Two-state European options in log-normal models were studied in Naik (1993); while European options in a two-state VG model were studied by Konikov and Madan (2000). Albanese, Jaimungal, and Rubisov (2003) derive closed form results for barrier and European options, and semi-closed form formulae for American options, in a special class of two-state VG models. Here, we demonstrate that the FST algorithm can easily incorporate path-dependent options, such as barrier and American options, with multiple regimes and multiple assets in computational time proportional to the number of regimes.

Let $\mathbb{K} := \{1, \ldots, K\}$ denote the possible hidden states of the world, and let $Z(t) \in \mathbb{K}$ denote the prevailing state of the world at time $t$. We will assume that $Z(t)$ is driven by a continuous time Markov chain with generator $A$, i.e. the transition probability from state $k$ at time $t_1$ to state $l$ at time $t_2$ is $P_{kl}^{t_1 t_2} := \mathbf{Q}(Z(t_2) = l | Z(t_1) = k) = (\exp\{(t_2 - t_1)\mathbf{A}\})_{kl}$. The real matrix $\mathbf{A}$ satisfies the usual requirements: $A_{ll} = -\sum_{k \neq l} A_{lk}$ and $A_{kl} \geq 0 \; \forall k \neq l$. Given that $Z(t) = k$, we assume that the joint stock price process $\mathbf{S}(t)$ follows a $d$-dimensional exponential Lévy process with Lévy triple $(\boldsymbol{\gamma}^{(k)}, \mathbf{C}^{(k)}, \boldsymbol{\nu}^{(k)})$. The drift vectors of each state are assumed prefixed at their risk-neutral levels of $\gamma_j^{(k)} = r - \Psi^{(k)}(-i\mathbf{1}_j)$, where $\Psi^{(k)}(\boldsymbol{\omega})$ denotes the characteristic exponent of the respective Lévy processes and $\mathbf{1}_j$ is the vector with zeroes everywhere except a single entry of 1 at dimension $j$. This modeling assumption can succinctly be written $d\mathbf{X}(t) = d\mathbf{X}^{(Z(t))}(t)$, where $\mathbf{X}^{(k)}(t)$ is the $k$-th $d$-dimensional Lévy process and the price processes are obtained by exponentiation component wise: $S_j(t) = S_j(0) \exp\{X_j(t)\}$. Chourdakis (2005) investigates the $d = 1$ version of this framework and derives the characteristic function of the terminal stock price.

The author calculates European option prices via FFT methods; however, then resorts to numerical integration for the valuation of path-dependent options. We take a slightly different approach, and make use of a generalization of the FST algorithm which allows path-dependent options based on the regime switching models to be valued efficiently.

Under the above assumptions, let $v(\mathbf{X}(t), Z(t), t)$ denote the discounted-adjusted and log-transformed price at time $t$ conditional on the state $Z(t)$ and spot levels $\mathbf{X}(t)$. It is not difficult to show that European option prices satisfy the following system of PIDEs:

$$\begin{cases} \partial_t + \left(A_{kk} + \mathcal{L}^{(k)}\right) v(\mathbf{x}, k, t) + \sum_{j \neq k} A_{jk} \, v(\mathbf{x}, j, t) &= 0 \,, \\ \qquad\qquad\qquad\qquad\qquad v(\mathbf{x}, k, T) &= \varphi(\mathbf{S}(0)e^{\mathbf{x}}) \,, \end{cases} \tag{41}$$

for every $k \in \mathbb{K}$. Here, $\mathcal{L}^{(k)}$ represents the infinitesimal generator of the $k$-th $d$-dimensional Lévy process. It is possible in principle to apply any of the usual finite-difference schemes to this system of PIDEs to solve the problem. However, as discussed earlier, this is quite difficult due to the non-local integral terms and especially so for multi-dimensional problems. Instead, we develop an FST algorithm. First discretize the continuous time Markov chain in the usual manner by partitioning time into steps of size $\Delta t$ and assuming $Z(t)$ is held constant on time

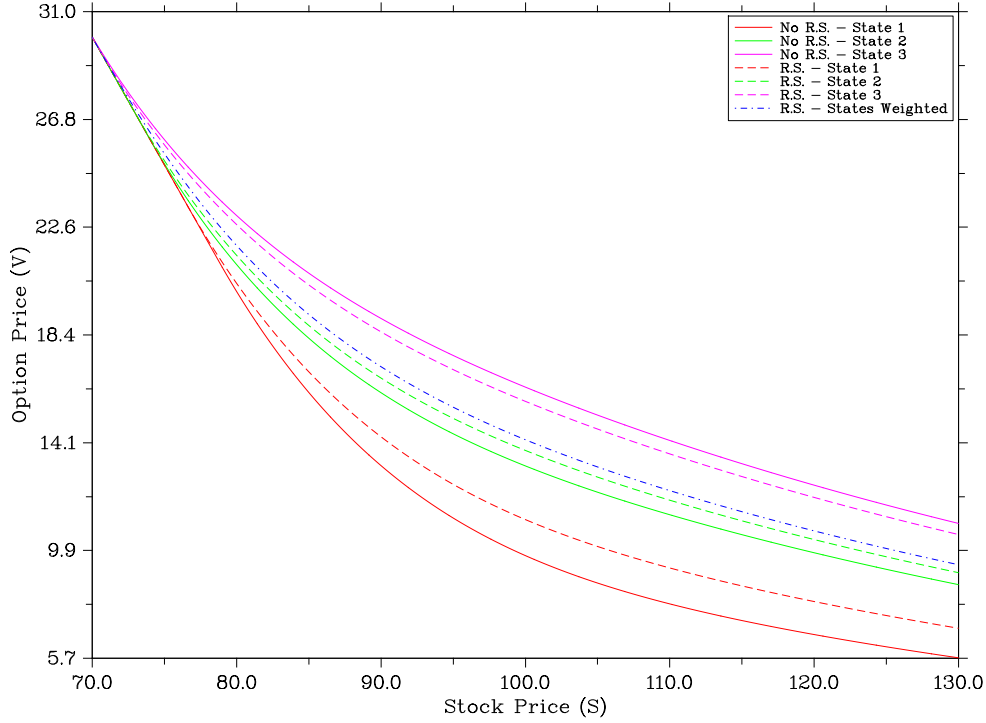

Figure 3. This diagram depicts the effect of introduction of regime switching on option price of an American option. *Option:* American put $S = 100.0, K = 100.0, T = 1.0$; *Model:* Merton jump-diffusion $\sigma = 0.15, \tilde{\mu} = -0.5, \tilde{\sigma} = 0.45, r = 0.05$ *Regime Switching:* parameter states $\lambda \in [0.3, 0.5, 0.7]$, initial state probabilities $q = [0.2, 0.3, 0.5]$, Markov chain generator $A = [-0.4, 0.3, 0.1; 0.1, -0.5, 0.4; 0.05, 0.15, -0.2]$

intervals $(t_{n-1}, t_n]$ (with $t_n = n\Delta t$), for $n \in \mathbb{N}$, with transition probabilities

$$P_{kl} := \begin{cases} A_{kl}\,\Delta t\,, & k \neq l\,, \\ 1 + A_{ll}\,\Delta t\,, & \text{otherwise}\,. \end{cases} \tag{42}$$

Then, by the martingale property of $v$ and the law of iterated expectations we have

$$\begin{aligned} v(\mathbf{X}(t_n), Z(t_n), t_n) &= \mathbb{E}_{t_n}^{\mathbb{Q}}\left[v(\mathbf{X}(t_{n+1}), Z(t_{n+1}), t_{n+1})\right] \\ &= \mathbb{E}_{t_n}^{\mathbb{Q}}\left[\mathbb{E}^{\mathbb{Q}}\left[v(\mathbf{X}(t_{n+1}), Z(t_{n+1}), t_{n+1})|\mathcal{F}_{t_n} \vee Z(t_{n+1})\right]\right]\,. \end{aligned}$$

Within the inner expectation, the process $\mathbf{X}(t)$ follows a given $d$-dimensional Lévy model, resulting in an expectation of the single regime form; consequently, the inner expectation can be written

$$\mathcal{F}^{-1}\left[\mathcal{F}[v](\boldsymbol{\omega}, Z(t_{n+1}), t_{n+1})\exp\left\{\Delta t\,\Psi^{(Z(t_{n+1}))}(\boldsymbol{\omega})\right\}\right](\mathbf{X}(t_n))\,. \tag{43}$$

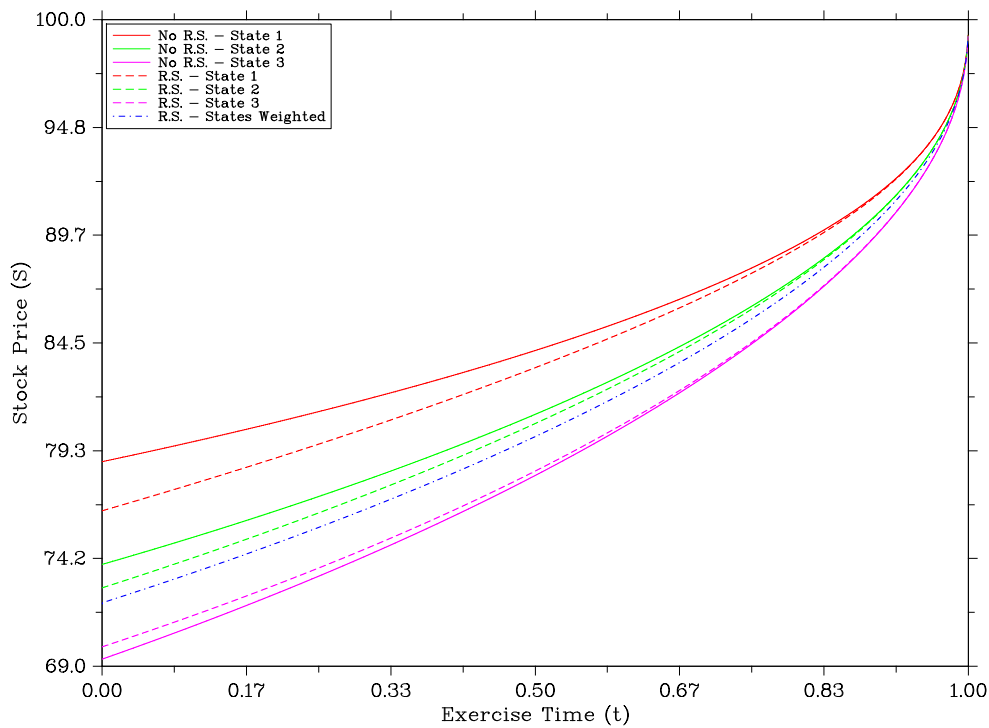

Figure 4. This diagram depicts the effect of introduction of regime switching on exercise boundary of an American option. *Option:* American put $S = 100.0, K = 100.0, T = 1.0$; *Model:* Merton jump-diffusion $\sigma = 0.15, \tilde{\mu} = -0.5, \tilde{\sigma} = 0.45, r = 0.05$ *Regime Switching:* parameter states $\lambda \in [0.3, 0.5, 0.7]$, initial state probabilities $q = [0.2, 0.3, 0.5]$, Markov chain generator $A = [-0.4, 0.3, 0.1; 0.1, -0.5, 0.4; 0.05, 0.15, -0.2]$

The above representation can be interpreted as the (discount-adjusted) price of a contingent claim with payoff $v(\mathbf{X}(t_{n+1}), Z(t_{n+1}), t_{n+1})$ assuming that the multi-dimensional price process is following $d$-dimensional Lévy model $Z(t_{n+1})$. Finally, due to the linearity of the inverse Fourier transform, the outer expectation can be computed to obtain the simple iterative scheme

$$v(\mathbf{X}(t_n), j, t_n) \;\; = \;\; \sum_{k=1}^{K} P_{jk} \, \mathcal{F}^{-1} \left[ \mathcal{F}[v](\boldsymbol{\omega}, k, t_{n+1}) \, \exp\left\{ \Delta t \, \Psi^{(k)}(\boldsymbol{\omega}) \right\} \right] (\mathbf{X}(t_n)) \, . \qquad (44)$$

At each time step, the algorithm therefore requires storing $K$ prices. These $K$ prices are then integrated backwards in time by the FST algorithm, then weighted according to the transition probabilities. If there are exercise decisions to be made, these must be made after averaging. In this manner, the price of the option today in all $K$ states will be known. Since a trader will likely not know for certain what state the world is currently in, the price will not simply be one of these prices; rather, the trader must decide, exogenously, on a probability $q_k$ that the

| N | M | Value | Change | $\log_2$Ratio | CPU-Time |
|---|---|---|---|---|---|
| 2048 | 128 | 33.76666279 | | | 0.104524 |
| 4096 | 512 | 33.77807492 | 0.011412 | | 0.693101 |
| 8192 | 2048 | 33.78118573 | 0.003111 | 1.8752 | 6.960128 |
| 16384 | 8192 | 33.78168754 | 0.000502 | 2.6321 | 64.977226 |
| 32768 | 32768 | 33.78181086 | 0.000123 | 2.0248 | 656.755101 |

Table 6

*Option:* American put $S = 1369.41, K = 1200.0, T = 0.56164$; *Model:* Variance-Gamma $\mu = -0.22898, \sigma = 0.20722, r = 0.0541, q = 0.012$; *Regime Switching:* parameter states $\kappa \in [0.22460, 0.49083, 0.50215]$, initial state probabilities $q = [0.2, 0.3, 0.5]$, Markov chain generator $A = [-0.4, 0.3, 0.1; 0.1, -0.5, 0.4; 0.05, 0.15, -0.2]$

world is in state $k$. Once these probabilities are determined, the trader's price for the option is $\sum_{k=1}^{K} q_k\, v(\mathbf{X}(0), k, 0)$.

In Table 6 we present the pricing results of an American put option with regime switching. The price of the option with regime switching (33.781) is lower than the price of same American option with $\kappa = 0.50215$ and without regime switching (35.524). This is expected since the former option switches between periods of low, medium and high subordinator volatility while the latter option is always in the state of high subordinator volatility.

## 6. Conclusions and Future Work

We introduced a new method for pricing European and path-dependent options when the underlying process(es) follows a regime switching Lévy process(es). The method treats the integral term and diffusion terms in the pricing PIDE symmetrically, is efficient, and accurate.

The numerical results attained by the FST method are quite impressive. We succeeded in reproducing the pricing results obtained by various authors. For European options, our algorithm is preferable to PIDE finite-difference based methods since it does not require a time-stepping procedure. Our algorithm is more appealing than the usual Fourier transform methods since the analytic expression for the Fourier transform of the option payoff is not required. Thus, the FST method is the method of choice for European options, especially with non-standard payoffs. We showed that Barrier options are priced accurately by comparing with analytical results in the BSM model. Furthermore, we compare our results for Barrier options under jump models with those of other authors and find excellent agreement. Additionally, we matched results obtained by other published methods for American options. This confirms that the FST is a robust method for pricing options with excellent precision properties. Moreover, the FST method provides a generic framework for option pricing under any stock price process, such as Brownian motion, jump-diffusion or exponential-Lévy. We also demonstrated how the FST method easily extends to the multi-dimensional case and can incorporate regime switching.

At this stage of development, the FST algorithm is first order in time. It should be noted, however, that second order methods in time typically involve a multi-step or iterative refinement of the solution at each time step. It is of great interest to investigate various approaches to improve the FST time order. For American options, this may involve an iterative approach. Barrier options are more challenging than American options due to the discontinuity in the option value created by the knock-out / knock-in provisions. In finite-difference methods, this problem is addressed by refining the mesh around the point of discontinuity. FFT algorithms, however, are designed for uniformly spaced points, making mesh refinement difficult to implement. Fortunately, FFTs using unequal spaced data (NFFT) have been developed recently. Potts, Steidl, and Tasche (2000) provide an overview of existing NDFT methods and develop a method of their own. An interesting area of research would be the use of NFFT algorithms in the FST method to improve its time-stepping order. Other areas of potential research include: computation of American exercise boundary, efficient computation of the Greeks, processes parameter calibration, and stochastic optimal control problems.

## A. Pricing Results

| N | Value | Change | $\log_2$Ratio | CPU-Time |
|---|---|---|---|---|
| 2048 | 0.04261423 | | | 0.002460 |
| 4096 | 0.04263998 | 0.000026 | | 0.005240 |
| 8192 | 0.04264641 | 0.000006 | 2.0018 | 0.009759 |
| 16384 | 0.04264801 | 0.000002 | 2.0010 | 0.019089 |
| 32768 | 0.04264841 | 0.000000 | 2.0011 | 0.038259 |

Table 7

*Option:* European call $S = 1.0, K = 1.0, T = 0.2$; *Model:* Kou jump-diffusion $\sigma = 0.2, \lambda = 0.2, p = 0.5, \eta_- = 3, \eta_+ = 2, r = 0.0$; *Integral Price:* 0.0426478 *Quoted Price:* 0.0426761 *Source:* Almendral and Oosterlee (2005)

| N | Value | Change | $\log_2$Ratio | CPU-Time |
|---|---|---|---|---|
| 2048 | 7.49983308 | | | 0.002743 |
| 4096 | 7.50061473 | 0.000782 | | 0.005478 |
| 8192 | 7.50091004 | 0.000295 | 1.4043 | 0.011147 |
| 16384 | 7.50098386 | 0.000074 | 2.0000 | 0.024923 |
| 32768 | 7.50100232 | 0.000018 | 2.0002 | 0.043864 |

Table 8

*Option:* European call $S = 100.0, K = 100.0, T = 0.46575$; *Model:* Variance-Gamma $\mu = -0.28113, \sigma = 0.19071, \kappa = 0.49083, r = 0.0549, q = 0.011$; *Integral Price:* 7.50100847 *Source:* Hirsa and Madan (2004)

| N | Value | Change | $\log_2$Ratio | CPU-Time |
|---|---|---|---|---|
| 2048 | 0.10295883 | | | 0.005580 |
| 4096 | 0.10296489 | 0.000006 | | 0.009415 |
| 8192 | 0.10296640 | 0.000002 | 2.0009 | 0.018672 |
| 16384 | 0.10296678 | 0.000000 | 2.0004 | 0.031994 |
| 32768 | 0.10296687 | 0.000000 | 2.0002 | 0.056374 |

Table 9

*Option:* European put $S = 1.0, K = 1.0, T = 1.0$; *Model:* CGMY $C = 1.0, G = 5.0, M = 5.0, Y = 0.5, r = 0.1$; *Integral Price:* 0.10296691 *Quoted Price:* 0.1029669 *Source:* Almendral and Oosterlee (2007)

| N | M | Value | Change | log$_2$Ratio | CPU-Time |
|---|---|---|---|---|---|
| 2048 | 128 | 3.23945333 | | | 0.020389 |
| 4096 | 512 | 3.24080513 | 0.001352 | | 0.169292 |
| 8192 | 2048 | 3.24114185 | 0.000337 | 2.0053 | 1.736765 |
| 16384 | 8192 | 3.24122597 | 0.000084 | 2.0011 | 14.570725 |
| 32768 | 32768 | 3.24124692 | 0.000021 | 2.0050 | 150.485790 |

Table 10

*Option:* American put $S = 100.0, K = 100.0, T = 0.25$; *Model:* Merton jump-diffusion $\sigma = 0.15, \lambda = 0.1, \tilde{\mu} = -0.9, \tilde{\sigma} = 0.45, r = 0.05$; *Quoted Price:* 3.2412435 *Source:* d'Halluin, Forsyth, and Labahn (2003)

| N | M | Value | Change | log$_2$Ratio | CPU-Time |
|---|---|---|---|---|---|
| 2048 | 128 | 35.50814417 | | | 0.021277 |
| 4096 | 512 | 35.51990755 | 0.011763 | | 0.162960 |
| 8192 | 2048 | 35.52355311 | 0.003646 | 1.6901 | 1.646470 |
| 16384 | 8192 | 35.52432105 | 0.000768 | 2.2471 | 15.200406 |
| 32768 | 32768 | 35.52449938 | 0.000178 | 2.1065 | 142.993274 |

Table 11

*Option:* American put $S = 1369.41, K = 1200.0, T = 0.56164$; *Model:* Variance-Gamma $\mu = -0.22898, \sigma = 0.20722, \kappa = 0.50215, r = 0.0541, q = 0.012$; *Quoted Price:* 35.5301 *Source:* Hirsa and Madan (2004)

| N | M | Value | Change | log$_2$Ratio | CPU-Time |
|---|---|---|---|---|---|
| 2048 | 128 | 8.89073604 | | | 0.026034 |
| 4096 | 512 | 8.89040553 | 0.000331 | | 0.188495 |
| 8192 | 2048 | 8.89031967 | 0.000086 | 1.9446 | 1.261334 |
| 16384 | 8192 | 8.89029787 | 0.000022 | 1.9781 | 10.623886 |
| 32768 | 32768 | 8.89029207 | 0.000006 | 1.9105 | 126.700997 |

Table 12

*Option:* Down-and-Out Barrier Call $S = 100.0, K = 110.0, B = 85, T = 1.0$; *Model:* Merton jump-diffusion $\sigma = 0.25, \lambda = 2.0, \tilde{\mu} = 0.0, \tilde{\sigma} = 0.1, r = 0.05$; *Quoted Price:* 9.013 *Source:* Metwally and Atiya (2003)

## B. Time-Stepping Results

| N | M | Value | Change | log$_2$Ratio | CPU-Time |
|---|---|-------|--------|-----------|----------|
| 4096 | 512 | 35.51990755 | | | 0.147402 |
| 4096 | 1024 | 35.52162452 | 0.001717 | | 0.495313 |
| 4096 | 2048 | 35.52246665 | 0.000842 | 1.0277 | 0.651053 |
| 4096 | 4096 | 35.52288380 | 0.000417 | 1.0135 | 1.427344 |
| 8192 | 512 | 35.52073332 | | | 0.403792 |
| 8192 | 1024 | 35.52261232 | 0.001879 | | 0.852099 |
| 8192 | 2048 | 35.52355311 | 0.000941 | 0.9980 | 1.612530 |
| 8192 | 4096 | 35.52402453 | 0.000471 | 0.9969 | 3.503661 |
| 16384 | 512 | 35.52079913 | | | 0.989980 |
| 16384 | 1024 | 35.52267988 | 0.001881 | | 1.735797 |
| 16384 | 2048 | 35.52361898 | 0.000939 | 1.0020 | 3.995848 |
| 16384 | 4096 | 35.52408731 | 0.000468 | 1.0038 | 7.307692 |

Table 13

*Option:* American put $S = 1369.41, K = 1200.0, T = 0.56164$; *Model:* Variance-Gamma $\mu = -0.22898, \sigma = 0.20722, \kappa = 0.50215, r = 0.0541, q = 0.012$; *Quoted Price:* 35.5301 *Source:* Hirsa and Madan (2004)

| N | M | Value | Change | log$_2$Ratio | CPU-Time |
|---|---|-------|--------|-----------|----------|
| 4096 | 512 | 9.22447378 | | | 0.347249 |
| 4096 | 1024 | 9.22493756 | 0.000464 | | 0.874211 |
| 4096 | 2048 | 9.22517293 | 0.000235 | 0.9785 | 1.002758 |
| 4096 | 4096 | 9.22529067 | 0.000118 | 0.9993 | 1.798958 |
| 8192 | 512 | 9.22451151 | | | 0.689149 |
| 8192 | 1024 | 9.22497177 | 0.000460 | | 1.288446 |
| 8192 | 2048 | 9.22520088 | 0.000229 | 1.0064 | 1.843247 |
| 8192 | 4096 | 9.22531447 | 0.000114 | 1.0122 | 4.063315 |
| 16384 | 512 | 9.22451347 | | | 1.172726 |
| 16384 | 1024 | 9.22497714 | 0.000464 | | 2.521093 |
| 16384 | 2048 | 9.22520900 | 0.000232 | 0.9999 | 4.401092 |
| 16384 | 4096 | 9.22532453 | 0.000116 | 1.0050 | 9.088098 |

Table 14

*Option:* American put $S = 90.0, K = 98.0, T = 0.25$; *Model:* CGMY $C = 0.42, G = 4.37, M = 191.2, Y = 1.0102, r = 0.1$; *Quoted Price:* 9.2185 *Source:* Forsyth, Wan, and Wang (2006)

| N | M | Value | Change | log$_2$Ratio | CPU-Time |
|---|---|---|---|---|---|
| 4096 | 512 | 0.25401054 | | | 0.168113 |
| 4096 | 1024 | 0.25409348 | 0.000083 | | 0.310622 |
| 4096 | 2048 | 0.25413559 | 0.000042 | 0.9779 | 0.717033 |
| 4096 | 4096 | 0.25415687 | 0.000021 | 0.9844 | 1.426751 |
| 8192 | 512 | 0.25402488 | | | 0.466672 |
| 8192 | 1024 | 0.25410782 | 0.000083 | | 0.985679 |
| 8192 | 2048 | 0.25414993 | 0.000042 | 0.9779 | 1.661753 |
| 8192 | 4096 | 0.25417121 | 0.000021 | 0.9844 | 3.377130 |
| 16384 | 512 | 0.25402651 | | | 1.070065 |
| 16384 | 1024 | 0.25410944 | 0.000083 | | 2.567016 |
| 16384 | 2048 | 0.25415155 | 0.000042 | 0.9779 | 4.697512 |
| 16384 | 4096 | 0.25417284 | 0.000021 | 0.9844 | 7.583904 |

Table 15

*Option:* Up-and-Out Barrier Call $S = 100.0, K = 100.0, B = 110.0, T = 1.0$; *Model:* Black-Scholes-Merton $\sigma = 0.15, r = 0.05, q = 0.02$; *Closed-Form Price:* 0.2541963 *Source:* Hull (2005)

| N | M | Value | Change | log$_2$Ratio | CPU-Time |
|---|---|---|---|---|---|
| 4096 | 512 | 8.89040553 | | | 0.170580 |
| 4096 | 1024 | 8.89035362 | 0.000052 | | 0.412546 |
| 4096 | 2048 | 8.89032768 | 0.000026 | 1.0006 | 0.890808 |
| 4096 | 4096 | 8.89031471 | 0.000013 | 1.0003 | 1.822582 |
| 8192 | 512 | 8.89039752 | | | 0.432691 |
| 8192 | 1024 | 8.89034561 | 0.000052 | | 1.296842 |
| 8192 | 2048 | 8.89031967 | 0.000026 | 1.0006 | 1.600545 |
| 8192 | 4096 | 8.89030670 | 0.000013 | 1.0003 | 3.302225 |
| 16384 | 512 | 8.89039517 | | | 1.257726 |
| 16384 | 1024 | 8.89034327 | 0.000052 | | 2.335755 |
| 16384 | 2048 | 8.89031732 | 0.000026 | 1.0005 | 4.781386 |
| 16384 | 4096 | 8.89030435 | 0.000013 | 1.0004 | 9.956966 |

Table 16

*Option:* Down-and-Out Barrier Call $S = 100.0, K = 110.0, B = 85.0, T = 1.0$; *Model:* Merton jump-diffusion $\sigma = 0.25, \lambda = 2.0, \tilde{\mu} = 0.0, \tilde{\sigma} = 0.1, r = 0.05$; *Quoted Price:* 9.013 *Source:* Metwally and Atiya (2003)

## C. Multi-Asset Pricing Results

| N | Value | Change | $\log_2$Ratio | CPU-Time |
|------|------------|----------|----------|------------|
| 512  | 7.55540525 |          |          | 0.450042   |
| 1024 | 7.53923270 | 0.016173 |          | 1.944840   |
| 2048 | 7.54214361 | 0.002911 | 2.4740   | 8.654331   |
| 4096 | 7.54233725 | 0.000194 | 3.9100   | 37.481155  |
| 8192 | 7.54232390 | 0.000013 | 3.8581   | 159.032330 |

Table 17

*Option:* Spread call $\alpha = 1.0, S_1 = 96.0, \beta = 1.0, S_2 = 100.0, K = 2.0, T = 1.0$; *Model:* Black-Scholes-Merton $\sigma_1 = 0.1, q_1 = 0.05, \sigma_2 = 0.2, q_2 = 0.05, \rho = 0.5, r = 0.1$; *Kirk's Formula Price:* 7.54232193 *Quoted Price:* 7.542242 *Source:* Dempster and Hong (2000)

| N | Value | Change | $\log_2$Ratio | CPU-Time |
|------|-------------|----------|----------|------------|
| 512  | 15.03639950 |          |          | 0.880245   |
| 1024 | 15.02776432 | 0.008635 |          | 2.821585   |
| 2048 | 15.02919574 | 0.001431 | 2.5928   | 11.919293  |
| 4096 | 15.02924971 | 0.000054 | 4.7293   | 48.371978  |
| 8192 | 15.02924214 | 0.000008 | 2.8345   | 209.806361 |

Table 18

*Option:* Spread call $S_1 = 96.0, S_2 = 100.0, K = 2.0, T = 1.0$; *Model:* Merton jump-diffusion $\sigma_1 = 0.1, q_1 = 0.05, \lambda_1 = 0.25, \bar{\mu_1} = -0.13, \bar{\sigma_1} = 0.37, \sigma_2 = 0.2, q_2 = 0.05, \lambda_2 = 0.5, \bar{\mu_2} = 0.11, \bar{\sigma_2} = 0.41, \rho = 0.5, r = 0.1$; *Kirk's Formula Price:* 15.03001533

| N | M | Value | Change | $\log_2$Ratio | CPU-Time |
|------|------|------------|----------|----------|--------------|
| 512  | 64   | 5.61098208 |          |          | 3.009590     |
| 1024 | 256  | 5.61273425 | 0.001752 |          | 51.875703    |
| 2048 | 1024 | 5.61216093 | 0.000573 | 1.6117   | 1373.959488  |
| 4096 | 4096 | 5.61199872 | 0.000162 | 1.8215   | 30498.836319 |

Table 19

*Option:* American spread put $\alpha = 1.0, S_1 = 96.0, \beta = 1.0, S_2 = 100.0, K = 2.0, T = 1.0$; *Model:* Black-Scholes-Merton $\sigma_1 = 0.1, q_1 = 0.05, \sigma_2 = 0.2, q_2 = 0.05, \rho = 0.5, r = 0.1$;